\documentclass{article}
\usepackage{amsmath}
\usepackage{latexsym,amssymb,amsthm,graphicx}
\newcommand\vv{\mathbf v} \newcommand\al{\alpha}
\newcommand\D{\mathrm D} \newcommand\oo{\mathrm o}
\newcommand\UU{\mathbf U} \newcommand\V{\mathbf V}
\newcommand\rr{\mathbb R}
\newcommand\xx{\mathbf x} \newcommand\Xx{\mathbf X}
\newcommand\ts{\mathbf t}
\newcommand\ttau{\boldsymbol\tau}
 \newcommand\cc{\mathbf c}
\newcommand\uu{\mathbf u} 
\newcommand\R{\mathbf R}
\newcommand\C{\mathbf C}
\newcommand\I{\mathbf I}
\newcommand\M{\mathbf M} \newcommand\W{\mathbf W}
\newcommand\w{\mathbf w}
\newcommand\K{\mathbf K} \newcommand\T{\mathbf T}
\newcommand\A{\mathbf A}
\newcommand\q{\mathbf q} \newcommand\dd{\mathrm d}
\newcommand\Om{\boldsymbol\Omega}\newcommand\kk{\mathbf k}
\newcommand\E{\mathbf E}\newcommand\om{\Om_{_\UU}}
\newcommand\nnab{\boldsymbol\nabla}\newcommand\ii{\mathbf i}

\newcommand\re[1]{(\ref{#1})}
\newcommand\ol[1]{\overline{#1}}
\newcommand\ul[1]{\underline{#1}}

\newcommand{\be}{\begin{equation}}
\newcommand{\ee}{\end{equation}}
\newcommand{\ben}{\begin{equation*}}
\newcommand{\een}{\end{equation*}}
\newcommand{\bea}{\begin{eqnarray}}
\newcommand{\eea}{\end{eqnarray}}
\newcommand{\eean}{\end{eqnarray*}}
\newcommand{\bean}{\begin{eqnarray*}}
\newcommand{\bma}{\begin{pmatrix}}
\newcommand{\ema}{\end{pmatrix}}

\begin{document}

\title{Absolute Time Derivatives}
\author{T. Matolcsi \\
Department of Applied Analysis, \\
E\"otv\"os Lor\'and University, Budapest, Hungary\\
P. V\'an \\
Department of Theoretical Physics\\
KFKI Research Institute for Particle and Nuclear Physics,
\\Budapest, Hungary }

\maketitle

\begin{abstract}
 A four dimensional treatment of nonrelativistic space-time
gives a natural frame to deal with objective time derivatives. In
this framework some well known objective time derivatives of
continuum mechanics appear as Lie-derivatives. Their coordinatized
forms depends on the tensorial properties of the relevant physical
quantities. We calculate the particular forms of objective time
derivatives for scalars, vectors, covectors and different second
order tensors from the point of view of a rotating observer. The
relation of substantial, material and objective time derivatives is
treated.
\end{abstract}

\section{Introduction}

Objectivity  plays a fundamental role in continuum physics. Its
usual definition is based on time-dependent Euclidean
transformations. Some problems arise from it which mainly concern
quantities containing derivatives; they take their origin from the
fact that objectivity is defined for three-dimensional vectors but
differentiation -- with respect to time and space together --
results in a four-dimensional covector. Using a four-dimensional
setting, we have extended the notion of objectivity \cite{MatVan06a}
which puts the objectivity of material time derivatives into new
light.

More closely, $\partial_0 + \vv\cdot\nabla$ is usually considered to
be material time derivation. This applied to scalars results in
scalars but applied to an objective vector does not result in an
objective vector; that is why it is usually stated that this
operation is not objective. One of the most important aspects of our
four-dimensional treatment is the existence of a covariant
derivation in nonrelativistic space-time which results in that the
correct form of material time derivation for vectors depends on the
observer. For a rotating observer the material time derivative is
$\partial_0 + \vv\cdot\nabla + \Om$ where $\Om$ is the angular
velocity (vorticity) of the observer.

From a mathematical point of view, $\Om$ is a component of the
four-dimensional Christoffel symbols corresponding to the observer.
In the usual three-dimensional treatment four-dimensional
Christoffel symbols cannot appear. As a consequence, one looks for
`objective time derivatives' in such a way that $\partial_0 +
\vv\cdot\nabla$ is supplemented by some terms for getting an
objective operation which does not involve Christoffel symbols and
contains only partial derivatives. This is how one obtains the
`lower convected time derivative', the `upper convected time
derivative' and the Jaumann or `corotational time derivative', as it
is written in several textbooks and monographs of continuum
mechanics (e.g. \cite{Hau00b,Ber05b,Ott05b}) and especially of
rheology (e.g. \cite{BirAta77b,Tan85b}). The corotational time
derivative was first introduced by Jaumann \cite{Jau11a}, and the
convected derivatives by Oldroyd \cite{Old49a}.

In the present paper we investigate these derivatives from a
four-dimensional point of view. For getting a convenient insight in
their physical meaning, we apply a coordinate-free formulation of
nonrelativistic space-time.

In the second section we shortly summarize the essentials of the
space-time model. In the third section we introduce the observers
and space-time splittings on the example of rigid observer. Then
continuous media is treated. A four dimensional version of the
material manifold is a general observer in our absolute framework.
At the fifth section we give the material time derivatives of the
physical quantities of different tensorial order. Finally a summary
and a discussion of the results follows.

\section{Fundamentals of nonrelativistic space-time \\
model}\label{st}

In this section some notions and results of the nonrelativistic
space-time model as a mathematical structure \cite{Mat84b,Mat93b}
will be recapitulated.

\subsection{The structure of nonrelativistic space-time model}\label{ss:fund}

A {\em nonrelativistic space-time model} consists of

-- the {\em space-time} $M$ which is a four-dimensional oriented
affine space over the vector space $\M$,

-- the {\em absolute time} $I$ which is a one-dimensional oriented
affine space over the vector space $\I$ ({\em  measure line of time periods}),

-- the {\em time evaluation} $\tau:M\to I$ which is an affine surjection
over the linear map $\ttau:\M\to\I$,

-- the {\em measure line of distances} $\mathbf D$ which is a one
dimensional oriented vector space,

-- the {\em Euclidean structure} $\cdot :\mathbf E\times\mathbf E\to
\mathbf D\otimes\mathbf D$ which is a positive definite symmetric bilinear map
where
\begin{equation*}
 \mathbf E:=\mathrm{Ker}\ttau\subset \M
\end{equation*}
is the (three-dimensional) linear subspace of {\em spacelike
vectors}.

\begin{figure}[ht]
\centering
\includegraphics[height=6cm]{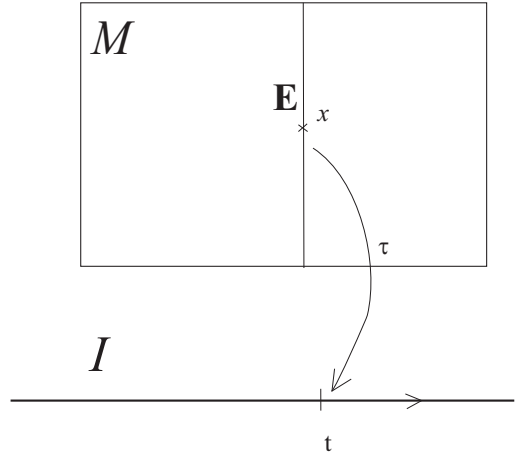}
\caption{Nonrelativistic space-time model} \label{f1}
\end{figure}

The time-lapse between the world points $x$ and $y$ is
$\tau(x)-\tau(y)=\ttau(x-y)$. Two world points are simultaneous if
the time-lapse between them is zero. The difference of two
simultaneous world points is a spacelike vector. The essential
elements of the model are visualized on figure \ref{f1}.

The length of the spacelike vector $\q$ is $|\q|:=\sqrt{\q\cdot\q}$.

The dual of $\M$, denoted by $\M^*$, is the vector space of
linear maps $\M\to\mathbb R$.  Elements of $\M^*$ are called
covectors.

In a similar way, the dual of $\E$ is $\E^*$.

If $\K\in\M^*$, i.e. $\K:\M\to\mathbb R$ is a linear map,
then its restriction to $\E$,  is an element of $\E^*$, denoted by $\K\cdot\ii$
which we call the absolute spacelike component of $\K$.

Note the important fact that the Euclidean structure allows us the
identification $\E^*\equiv\frac{\E}{\mathbf D\otimes\mathbf D}$. On the other
hand, {\em no similar identification is possible for} $\M^*$ because
there is no Euclidean or pseudo-Euclidean structure in $\M$. (In
coordinates: an element $\q$ of $\E$ is coordinatized as $q^i$ for
$i=1,2,3$ and $q_i=q^i$ can be written. On the other hand, an
element $\xx$ of $\M$, $\xx\notin\E$ is coordinatized as $x^\al$ for
$\al=0,1,2,3$ and $x_\al$ is not meaningful. Moreover, an element
$\K$ of $\M^*$ is coordinatized as $K_\al$ for $\al=0,1,2,3$ and
$K^\al=K_\al$ can be written for $\al=1,2,3$ but $K^0$ is not
meaningful.)

These features of vectors and covectors are consequences of the fact
that time is not embedded in space-time. This property of the model
eliminates such unphysical possibilities as the "angle" of space and
time. As a result the careful distinction of space-time vectors and
covectors is essential: there is no canonical way to identify them.

\subsection{Differentiation}

The affine structure of space-time implies the existence of an
absolute differentiation (in the language of manifolds: a
distinguished covariant differentiation).

If $V$ is a finite dimensional affine space over the vector space
$\V$, then a map
$A:M\to V$ is differentiable at $x$ if there is a linear map
$(\D A)(x):\M\to\V$ -- the derivative of $A$ at $x$ -- such that
\begin{equation*}
\lim_{y\to x}\frac{A(y)-A(x) - (\D A)(x)(y-x)}{||y-x||}=0
\end{equation*}
where $|| \ \ ||$ is an arbitrary norm on $\M$ (all norms on the
finite dimensional vector space $\M$ are equivalent).

As a consequence of the structure of our space-time model, the
partial time derivative of $A:M\to V$ makes no sense. On the other
hand, the {\em spacelike derivative} of $A$ is meaningful because
the spacelike vectors form a linear subspace in $\M$: $(\nabla
A)(x)$ is the derivative of the function $\E\to V$, $\q\mapsto
A(x+\q)$ at zero. It is evident then that $(\nabla A)(x)$ is the
restriction of the linear map $(\D A)(x)$ onto $\E$. The transpose
of a linear map $\mathbf A:\M\to\V$ is the linear map $\mathbf
A^*:\V^*\to \M^*$ defined by $\mathbf A^*\mathbf w:=\mathbf
w\circ\mathbf A$ for $\mathbf w\in\V^*$. Then, using the customary
identification of linear maps as tensors, we can consider
\begin{equation*}
\D A(x)\in\V\otimes\M^*, \qquad (\D A)^*(x)\in \M^*\otimes\V
\end{equation*}
and
\begin{equation*}
\xx\cdot(\D A)^*(x):= \D A(x) \xx\in\V, \qquad (\D A)^*(x)\mathbf
w\in \M^*
\end{equation*}
for $\xx\in\M$ and $\mathbf w\in\V^*$.

Accordingly,
\begin{equation}\label{nabla}
(\nabla A)(x)\in \V\otimes\E^*, \qquad (\nabla A)^*(x)\in
\E^*\otimes\V,
\end{equation}
\begin{equation*}
\q\cdot(\nabla A)^*(x):= (\nabla A)(x)\q \in\V, \qquad (\nabla
A)^*(x)\mathbf w\in \E^*
\end{equation*}
for $\q\in\E$ and $\mathbf w\in\V^*$.

In particular,

\begin{itemize}
\item the derivative of a scalar field $f:M\to\rr$ is a covector field,
$\D f(x)\in\M^*$,

$\bullet$ \ its spacelike derivative is a spacelike covector field,
$\nabla f(x)\in\E^*$;

\item the derivative of a vector field $\C:M\to\M$  is a mixed tensor
field, $(\D\C)(x)\in\M\otimes\M^*$ whose transpose is
$(\D\C)^*(x)\in\M^*\otimes\M$,

$\bullet$  \ its spacelike derivative is a mixed tensor field,
$(\nabla\C)(x)\in\M\otimes\E^*$ whose transpose is
$(\nabla\C)^*(x)\in\E^*\otimes\M$,

\item the spacelike derivative of a spacelike vector field $\cc:M\to\E$  is a
mixed spacelike tensor field, $(\nabla\cc)(x)\in\E\otimes\E^*$ whose
transpose is $(\nabla\cc)^*(x)\in\E^*\otimes\E$.

\item the derivative of a covector field $\K:M\to\M^*$  is a cotensor
field, $(\D\K)(x)\in\M^*\otimes\M^*$ whose transpose is
$(\D\K)^*(x)\in\M^*\otimes\M^*$.

\end{itemize}

Note that both the derivative of a covector field and its transpose are
in $\M^*\otimes\M^*$. Thus, we can define the {\em antisymmetric
derivative} of $\K$,
\begin{equation*}
(\D\land\K)(x):=(\D\K)^*(x) -(\D\K)(x).
\end{equation*}

On the contrary, the antisymmetric derivative of a vector field $\C:M\to\M$,
in general, does not make sense. The antisymmetric spacelike derivative of a
spacelike vector field $\cc:M\to\E^*$, however, can be defined
because the identification  $\E^*\equiv\frac{\E}{\mathbf D\otimes\mathbf D}$
implies
$\E\otimes\E^*\equiv\E^*\otimes\E$, so we can put
\begin{equation*}
(\nabla\land\cc)(x):=(\nabla\cc)^*(x) -(\nabla\cc)(x).
\end{equation*}

\section{Observers}

\subsection{Absolute velocity}

The history of a classical masspoint is described by
a {\em world line function}, a twice continuously
differentiable function $r:I\to M$ such that $\tau(r(t))=t$ for
all $t\in I$. A {\em world line} is the range of a
world line function; a world line is a curve in $M$.

If $r$ is a world line function, then $\ttau(\dot r(t))=1$. That is
why we call the elements of the set
\begin{equation*}
V(1):=\left\{\uu\in\frac{\M}{\I}\biggm|
\boldsymbol\tau(\uu)=1\right\}
\end{equation*}
{\em absolute velocities}. $V(1)$ is
a three dimensional affine space over $\frac{\E}{\I}$.

\subsection{Rigid observers}\label{rigid}

An observer, from a physical point of view, is a `continuous set
of material points'. Such a `continuous body' can be characterized
by assigning to any world point the absolute velocity of the particle
at that point, i.e. by an absolute velocity field. Thus  we accept that
an {\em observer} is a smooth map
\begin{equation*}
\UU:M\to V(1).
\end{equation*}

The integral curves of $\UU$ are world lines, representing the
histories of the material points that the observer is constituted
of. Thus it is quite evident that a maximal integral curve of $\UU$
is a {\em space point of the observer}. The set of the maximal
integral curves is the {\em space} of the observer, briefly the
$\UU$-{\em space}.

Keep in mind the most important -- but trivial -- fact concerning
observers: {\em a space point of an observer is a curve in
space-time}.
\begin{figure}[ht]
\centering
\includegraphics[height=6cm]{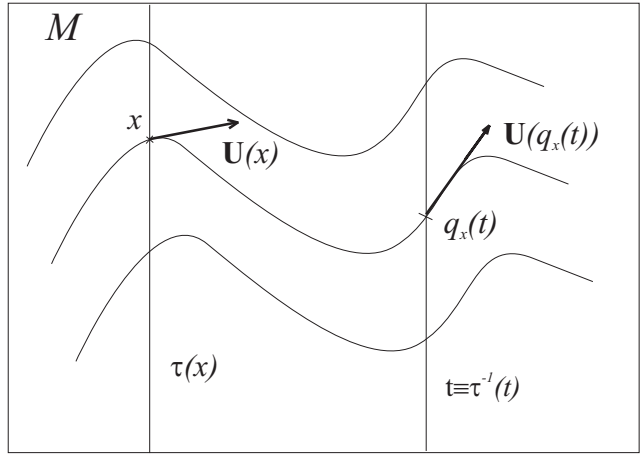}
\caption{Observers and observer spaces} \label{f2}
\end{figure}

Observers and their spaces are well defined simple and
straightforward notions. {\em The spaces of different observers
are evidently different}.

For an observer $\UU$, we denote by $q_x$ the world line function
whose range is the $\UU$-space point containing the world point $x$, i.e.
\begin{equation*}
\frac{\dd q_x(t)}{\dd t}= \UU(q_x(t)),\qquad
 q_x(\tau(x)) = x.
\end{equation*}

$\UU$ is {\em rigid} if the distance of any two
of its space points is time independent: given $x,y$ arbitrarily, then
 $|q_x(t)-q_y(t)| = |q_x(s)-q_y(s)|$ for all instants $t,s$.

It can be shown (\cite{Mat93b}, Chapter I.4) that the observer $\UU$
is rigid if and only if for all $t,t_\oo\in I$ there is a rotation
$\R(t,t_\oo)$ in $\E$ such that
\begin{equation}\label{forgas}
q_{x_\oo+ \q}(t)-q_{x_\oo}(t) = \R(t,t_\oo)\q \qquad
(\tau(x_\oo)=t_\oo,\q\in\E).
\end{equation}

Then putting $\dot\R(t,t_\oo):=\frac{\partial\R(t,t_\oo)}{\partial t}$,
\begin{equation*}
\Om(t):=\dot\R(t,t_\oo)\R(t,t_\oo)^{-1}\in \frac{\E\otimes\E^*}{\I}
\end{equation*}
is independent of $t_\oo$; it is the {\em angular velocity of the
rigid observer} at the instant $t$. It is easy to show that $\Om(t)$
is antisymmetric, moreover,
\begin{equation}\label{angvel}
\UU(x+\q)-\UU(x) = \Om(\tau(x))\q \qquad (x\in M, \q\in\E),
\end{equation}
which implies that $\nabla\UU(x)=\Om(\tau(x))$: {\em the spacelike
derivative of the rigid observer is its angular velocity
which is a spacelike antisymmetric tensor}.

Since the spacelike derivative of $\UU$  is antisymmetric, we have
$\nabla\UU(x)=-\frac1{2}(\nabla\land\UU)(x)$. This supports that later
\eqref{gangvel} is considered as the angular velocity of
an arbitrary (non-necessarily rigid) continuum.

An important particular rigid observer is the {\em inertial
observer}, when
 \begin{equation*}
 \UU(x) =  \text{const}.
\end{equation*}

\noindent therefore $\R$ is the identity of $\E$ and $\nabla\UU(x) = {\bf 0}$.

\subsection{Splitting of space-time by rigid observers}\label{ss:obsplit}

Let us consider an observer $\UU$. For every world point $x$ there
is a unique $\UU$-space point (world line, representing a point of
the observer) containing $x$ (the range of the world line function
$q_x$). Accordingly, the observer perceives the world point $x$ as a
couple of its absolute instant $\tau(x)$ and the corresponding
$\UU$-space point. We say that the observer {\em splits space-time}
into the Cartesian product of time and $\UU$-space.

Since $\UU$-space is not a simple mathematical object, in general,
the splitting of space-time by $\UU$ is not simple either. To
overcome this uneasiness, we consider {\em vectorized splittings} in
which $\UU$-space is represented by $\E$ as follows.

Let $\UU$ be a rigid observer and let $o$ be a world point,
conceived as a chosen `origin' in space-time. Then a space point of
the observer will be represented by the spacelike vector which is
the difference between $o$ and the simultaneous world point of the
space point in question. More closely, the $\UU$-space point (world
line) containing the world point $x$ will be represented by
$q_x(\tau(o))-o$. To get explicitly how $q_x(\tau(o))-o$ depends on
$x$, let us put $x_\oo:=o$, $\q:=q_x(\tau(o))-o$ and $t=\tau(x)$ in
\eqref{forgas} (then $\tau(o)=t_\oo$) and take into account that
$q_{q_x(t_\oo)}(\tau(x))=x$; in this way we obtain the vectorized
splitting in the form
\begin{equation}\label{split}
H:M\to I\times \E,\quad x\mapsto \bigl(\tau(x),
\R(\tau(x))^{-1}\bigl(x - q_o(\tau(x))\bigr).
\end{equation}
Here and in the sequel, for the sake of brevity, $\R(t):=
\R(t,t_\oo)$.

\begin{figure}[ht]
\centering
\includegraphics[height=8cm]{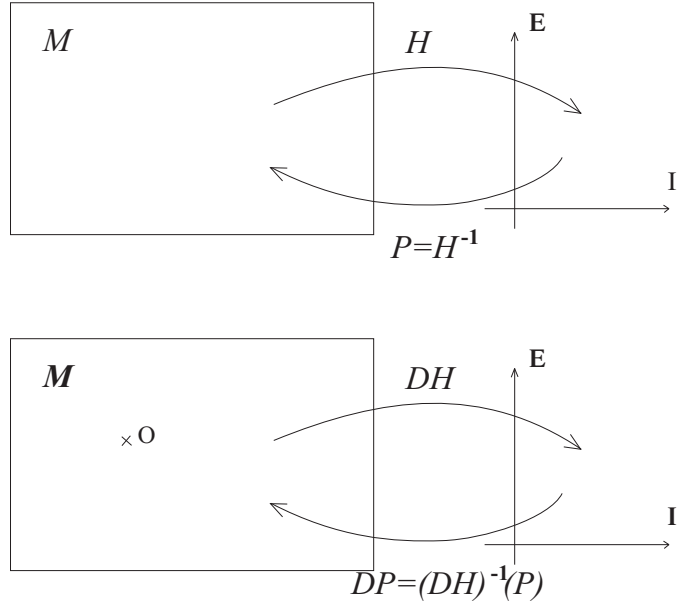}
\caption{Splitting of space-time} \label{f3}
\end{figure}

The observer splits $\M$, too, by the derivative of this space-time
splitting.

Differentiating $\R(\tau(x))^{-1}$ by $x$ we get
$-\R(\tau(x))^{-1}\dot\R(\tau(x))\R(\tau(x))^{-1}\ttau$, taking into
account $\dot q_o(\tau(x))=\UU(q_o(\tau(x))$ and the basic
properties of world line functions we find that the vectorized
splitting has the derivative
\begin{equation}\label{hder}
\D H(x)=\begin{pmatrix}\ttau \\ \R(\tau(x))^{-1}\left(\boldsymbol 1 -
\UU(x)\otimes\ttau\right)\end{pmatrix}:\M\to\I\times\E
\end{equation}
where $\boldsymbol 1$ is the identity of $\M$.

Note that restricting $\D H(x)$ onto $\E$, we obtain $\nabla H(x) =
(0, \R(\tau(x))^{-1})$; further we omit the zero component, thus we
consider that
\begin{equation*}
\nabla H(x) = \R(\tau(x))^{-1}:\E\to\E.
\end{equation*}

The inverse of the splitting is
\begin{equation}\label{invsplit}
H^{-1}=:P:I\times\E\to M, \quad (t,\q)\mapsto q_o(t) + \R(t)\q,
\end{equation}
whose partial derivatives are obtained easily:
\begin{equation}\label{pder0}
\partial_0 P(t,\q):=\frac{\partial P(t,\q)}{\partial t}=
\UU(q_o(t)) + \dot\R(t)\q = \UU(P(t,\q)),
\end{equation}
\begin{equation}\label{pder1}
\nnab P(t,\q):=\frac{\partial P(t,\q)}{\partial \q} =\R(t).
\end{equation}

The derivative of $P$ is the couple of the partial derivatives.
Differentiating the equality $H(P(t,\q))=(t,\q)$, we deduce
\begin{equation}\label{dhp}
\biggl(\partial_0 P, \nnab P\biggr)=(\D H(P))^{-1}.
\end{equation}

\subsection{Relative form of absolute physical
quantities}\label{splisec}

Using the splitting of $M$ and $\M$, a rigid observer $\UU$
represents physical quantities -- functions defined in space-time --
as functions defined in time and $\UU$-space. The splitting of the
space-time functions gives their $\UU$-{\em relative form}, the
usual field quantities defined on time and space.

The $\UU$-relative form of a \emph{scalar field} $f:M\to\rr$ is
\begin{equation}\label{scalsp}
f_{_\UU}:I\times\E\to\rr, \quad (t,\q)\mapsto f(P(t,\q)),
\end{equation}
briefly: $f_{_\UU}=f(P)$.

The $\UU$-relative form of a \emph{vector field} $\C:M\to\M$ is
\begin{equation*}
\C_{_\UU}:I\times\E\to\I\times\E, \quad (t,\q)\mapsto
\D H(P(t,\q))\C(P(t,\q)).
\end{equation*}
Using \eqref{hder} and an abbreviated notation, we have
\begin{equation}\label{vecsp}
\C_{_\UU} = \begin{pmatrix}\ttau\C(P) \\
\R^{-1}\bigl(\C(P) - \UU(P)\ttau\C(P)\bigr)\end{pmatrix}
\end{equation}

In particular, a \emph{spacelike vector field} $\cc:M\to \E$ has the
$\UU$-relative form (the trivial zero component omitted)
\begin{equation}\label{slvecsp}
\cc_{_\UU} =\R^{-1}\cc(P).
\end{equation}

Similarly, a \emph{spacelike tensor field} $\mathbf f:M\to
\E\otimes\E^*$ has the $\UU$-relative form
\begin{equation}\label{sltensp}
\mathbf f_{_\UU}= \R^{-1}\mathbf f(P)\R.
\end{equation}

The $\UU$-relative form of a \emph{covector field} $\K:M\to\M^*$ is
\begin{equation*}
\K_{_\UU}:= \left((\D H(P))^{-1}\right)^*\K(P)=\K(P)(\D H(P))^{-1}:
I\times\E\to\I^*\times\E^*;
\end{equation*}
by \eqref{dhp}, \eqref{pder0} and \eqref{pder1}, we find
\begin{equation}\label{covecsp}
\K_{_\UU}=\K(P)\cdot\biggl(\partial_0 P, \nnab P\biggr)=
\biggl(\K(P)\cdot\UU(P), (\K(P)\cdot\ii)\R\biggr)
\end{equation}
(recall that $\K\cdot\ii$ denotes the absolute spacelike component of $\K$,
the restriction of $\K$ onto $\E$). Note that the spacelike component
can be written in the form $\R^{-1}(\K(P)\cdot\ii)$, too, because for an
orthogonal map we have $\R^*=\R^{-1}$.

As a consequence one may calculate the $\UU$-relative form of second
order tensors easily. For example, a \emph{mixed tensor field}
$\mathbf F: M\to \E\otimes\M^*$ has the $\UU$-relative form
\begin{equation}\label{comixsp}
\mathbf F_{_\UU} = \biggl(\R^{-1}\mathbf F(P)\cdot\UU(P) ,
\R^{-1}(\mathbf F(P)\cdot\ii)\R\biggr).
\end{equation}

\subsection{Relative form of absolute derivatives}

The \emph{derivative $\D f$ of a scalar field} $f$ is a covector
field, thus its $\UU$-relative form is
\begin{equation}\label{dfsp}
(\D f)_{_\UU}=  \D f(P)\cdot \biggl(\partial_0 P, \nnab P\biggr)=
\biggl(\partial_0 f_{_\UU}, \nnab f_{_\UU}\biggr).
\end{equation}

The \emph{derivative $\D\cc$ of a spacelike vector field} $\cc$ is a
mixed tensor field, so differentiating \eqref{slvecsp} and applying
\eqref{comixsp}, we get
\begin{equation}\label{dcsp}
(\D\cc)_{_\UU} =\biggl((\partial_0 + \om)\cc_{_\UU},
 \nnab \cc_{_\UU}\biggr)
\end{equation}
where
\begin{equation}
\om:= \R^{-1}\Om\R=\R^{-1}\dot\R
\end{equation}
is the relative form of the angular velocity of the observer.

As a consequence,
\begin{equation}
(\nabla\cc)_{_\UU} = \nnab\cc_{_\UU}.
\end{equation}

\section{Continuous media}

A continuum, from a physical point of view, is a `continuous set
of material points'. The history of such a `continuous body' can be
described by an absolute velocity field $\uu:M\to V(1)$ which is
supposed to be twice differentiable.

Note that both an observer and a continuum are given by an absolute
velocity field. Keep in mind that majuscule $\UU$ will refer to an observer
(an `observing body'), minuscule $\uu$ will refer to a continuum
(a `body to be observed'). An observer is mostly supposed to be rigid,
a continuum is never rigid. An observer has no other property besides
its velocity field, a continuum has other characteristics, too: density,
stress, temperature, etc.

\subsection{Velocity fields}

Recall that $V(1)$ is an affine space over $\frac{\E}{\I}$, thus
\begin{itemize}
    \item the derivative of an absolute velocity field $\uu:M\to
V(1)$ is a mixed tensor field,
$(\D\uu)(x)\in\frac{\E}{\I}\otimes\M^*$ having the transpose
$(\D\uu)^*(x)\in\M^*\otimes\frac{\E}{\I}$.
    \item the spacelike derivative of $\uu$ is a mixed spacelike tensor
field, $(\nabla\uu)(x)\in\frac{\E}{\I}\otimes\E^*$ having the
transpose $(\nabla\uu)^*(x)\in\E^*\otimes\frac{\E}{\I}$.
\end{itemize}

In view of the identification $\E^*\equiv\frac{\E}{\mathbf D\otimes\mathbf D}$,
both $(\nabla\uu)(x)$ and $(\nabla\uu)^*(x)$ are considered to be in
$\frac{\E\otimes\E}{\I\otimes\mathbf D\otimes\mathbf D}$, thus the antisymmetric
spacelike derivative of $\uu$ makes sense, too:
\begin{equation*}
(\nabla\land\uu)(x):=(\nabla\uu)^*(x) -(\nabla\uu)(x).
\end{equation*}

According to the end of Subsection \ref{rigid}, we can interpret
\begin{equation}\label{gangvel}
-\frac1{2}(\nabla\land\uu)(x)
\end{equation}
as the {\em angular velocity} (vorticity) of the continuum at the
world point $x$.

Now let us consider a rigid observer $\UU$ which `observes' the continuum
$\uu$. We deduce from \eqref{vecsp} that the $\UU$-relative form of $\uu$ is
\begin{equation}\label{velsp}
\uu_{_\UU}(t,\q)= \begin{pmatrix} 1 \\
\vv_{_\UU}(t,\q)\end{pmatrix}
\end{equation}
where
\begin{equation}
\vv_{_\UU}=\R^{-1}\bigl(\uu(P) - \UU(P)\bigr)
\label{urelu}
\end{equation}
is the $\UU$-relative velocity field.

Then we derive that
\begin{equation}
\nnab\vv_{_\UU} = \R^{-1}\bigl((\nabla\uu)(P)\nnab P -
(\nabla\UU)(P)\nnab P\bigr)
\end{equation}
from which, taking into account that $\nabla\uu$ is a spacelike
tensor and using \eqref{pder1} and \eqref{sltensp}, we have
\begin{equation}\label{nabu}
(\nabla\uu)_{_\UU} =  \nnab \vv_{_\UU} + \om\quad\mbox{or}\quad
(\nabla\uu)^*_{_\UU} =  (\nnab\vv_{_\UU})^* - \om.
\end{equation}

\subsection{The flow of a continuum}

A velocity field $\uu$, by the solution of differential equation
$\dot x=\uu(x)$, generates a {\em flow}, the map
\begin{equation*}
\I\times M\to M, \quad (\ts,x)\mapsto \Upsilon_\ts(x)
\end{equation*}
such that
\begin{equation}
\frac{\dd \Upsilon_\ts(x)}{\dd\ts}=\uu(\Upsilon_\ts(x)),\qquad
\Upsilon_0(x)=x. \label{flowdef}\end{equation}

Thus, $t\mapsto \Upsilon_{t-\tau(x)}(x)$ is a world line function of
$\uu$, describing the history of a particle of the continuum.

It is well known from the theory of differential equations
\cite{Arn92b} that for any fixed $\ts$ the map $M\to M$, $x\mapsto
\Upsilon_\ts(x)$ is a twice differentiable bijection whose inverse
is twice differentiable, too (it is a diffeomorphism). Consequently,
its derivative $\D \Upsilon_\ts(x):= \frac{\dd \Upsilon_\ts(x)}{\dd
x}$  is a linear bijection $\M\to\M$ (an element of
$\M\otimes\M^*$). Note that $\D \Upsilon_0(x)$ is the identity of
$\M$.

The order of differentiations can be interchanged, thus
\begin{equation}\label{dexc}
\frac{\dd\D \Upsilon_\ts(x)}{\dd\ts}\left.\right|_{\ts=0} =
\D\uu(x).
\end{equation}

The customary notions regarding the kinematics of the continuum are
connected to the {\bf U}-relative form of the flow. According to the
previous general case a rigid observer splits the flow of the
continuum into the {\em duration} {\bf t} of the motion, the time
elapsed from the initial instant, and the {\em motion function}
$\chi_\ts$ \cite{TruNol65b}, the relative position of the particles
of the continuum in the space of the rigid observer:
\begin{equation*}
 (\Upsilon_\ts)_{_\UU}: I\times \E \to \I\times \E, \quad
 (t,\Xx)\mapsto DH(P(t,\Xx))\Upsilon_\ts(P(t,\Xx)).
\end{equation*}

With the customary and shortened notation  one can get
\begin{equation*}
    (\Upsilon_\ts)_{_\UU}(t,\Xx)=
(\ttau(\Upsilon_\ts),\R^{-1}(\Upsilon_\ts - \UU \ttau(\Upsilon_\ts))
) =: (\ts, \chi_\ts(\Xx)).
\end{equation*}

The spacelike part of the domain of the $\UU$-relative flow
is called {\em reference configuration}, because $\chi_{\bf
0}(\Xx)=\Xx$ as a consequence of the second formula of \re{flowdef}.
Let us note that the spacelike component of the flow, the
motion function, is a relative notion, depends on the observer
\cite{BerSve01a}. Similarly, the usual concepts of {\em body} and
{\em material manifold} of continuum physics (see e.g.
\cite{TruNol65b,Mau99b}) are relative, too.

\section{Time derivatives}

In this section we consider a continuum having the absolute velocity
field $\uu$.

\subsection{Material time derivative}

Let a physical quantity be described by $A:M\to V$ where $V$ is a
finite dimensional affine space. The function $\ts\mapsto
A(\Upsilon_\ts(x))$ is the change in time of the quantity along an
integral curve i.e. at a particle of the continuum. We have by the
chain rule that
\begin{equation*}
\frac{\dd A(\Upsilon_\ts(x))}{\dd\ts}\left.\right|_{\ts=0} = \D
A(x)\cdot\uu(x) =\uu(x)\cdot(\D A)^*(x)=:(\D_\uu A)(x).
\end{equation*}

It is a matter of course that we call $\D_\uu A=(\D A)\cdot\uu$
the {\em material time derivative} of $A$ with respect to $\uu$.
Clearly, this is an absolute object, not depending on any observer.

The $\UU$-relative form of the material time derivative of a scalar
field $f:M\to\rr$ is obtained by \eqref{dfsp} and \eqref{velsp}:
\begin{align*}
(\D_\uu f)_{_\UU} =& (\D f\cdot\uu)_{_\UU}= (\D f)_{_\UU}\cdot(\uu)_{_\UU}= \\
=& (\partial_0 + \vv_{_\UU}\cdot\nnab)f_{_\UU}.
\end{align*}

The $\UU$-relative form of the material time derivative of a spacelike
vector field $\cc:M\to\E$ is obtained by \eqref{dcsp} and \eqref{velsp}:
\begin{align}\label{mattime}
(\D_\uu\cc)_{_\UU} =& \bigl(\D\cc\cdot\uu\bigr)_{_\UU}=
(\D\cc)_{_\UU}\cdot(\uu)_{_\UU} = \nonumber\\
=& \left(\partial_0 + \om +  \vv_{_\UU}\cdot\boldsymbol\nabla\right)
\cc_{_\UU}.
\end{align}

We emphasize that {\em material time differentiation is absolute}
(objective), does not depend on any observer and its correct
relative form by a rigid observer for absolute spacelike vector
fields is $\partial_0 +\om+ \vv_{_\UU}\cdot\nnab$. We repeat for a
clear distinction: {\em the non-objective $\partial_0 +
\vv_{_\UU}\cdot\nnab$ is not the relative form of the material time
differentiation for spacelike vector fields} \cite{MatVan06a}.

\subsection{Traditional convected time derivatives}

\subsubsection{Upper convected time derivative}

Now we have to make a remark. Let $N$ be an affine space and let
$H:M\to N$ be a diffeomorphism. Then the vector field $\C:M\to\M$ is
sent by $H$ to the vector field $N\to\mathbf N$, $y\mapsto\D
H(y)\C(H^{-1}(y))$. This formula is applied when defining the split
form \eqref{vecsp} of a vector field and offers itself for the flow
generated by the velocity field, $H$ replaced with
$\Upsilon_\ts^{-1}$.

\begin{figure}[ht]
\centering
\includegraphics[height=6cm]{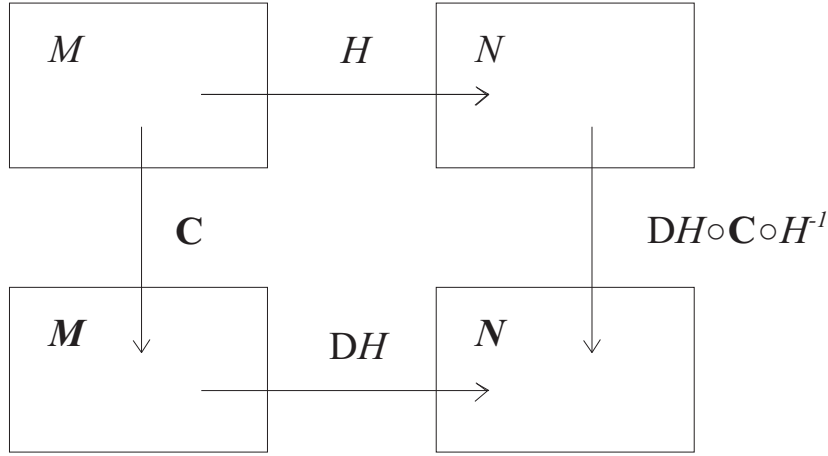}
\caption{Pull-back of a vector to the material manifold} \label{f4}
\end{figure}

Thus, instead of $t\mapsto \C(\Upsilon_\ts(x))$, it seems preferable
to consider \newline $\ts\mapsto (\D
\Upsilon_\ts(x))^{-1}\C(\Upsilon_\ts(x))$ as the change in time of
the vector field along a particle of the continuum. Since
$$
 \frac{\dd(\D \Upsilon_\ts(x))^{-1}}{\dd\ts} = -(\D
\Upsilon_\ts(x))^{-1}\frac{\dd\D \Upsilon_\ts(x)}{\dd\ts}(\D
\Upsilon_\ts(x))^{-1},
$$
so
\begin{equation}
\left.\frac{\dd(\D\Upsilon_\ts(x))^{-1}\C(\Upsilon_\ts(x))}
{\dd\ts}\right|_{\ts=0} =
\uu(x)\cdot(\D\C)^*(x)-\C(x)\cdot(\D\uu)^*(x) =:(L_\uu\C)(x).
\label{Lvect}\end{equation}

We mention that $L_\uu\C$ is known in differential geometry as the
Lie derivative of $\C$ by $\uu$ \cite{ChoAta82b}.

The first term of the Lie derivative is just the material time
derivative.

For a spacelike vector field $\cc:M\to\E$ we have
\begin{equation}\label{Lderv}
L_\uu\cc= \D_\uu\cc - \cc\cdot(\nabla\uu)^*.
\end{equation}

The $\UU$-relative form of the Lie derivative of the spacelike
vector field $\cc$ is
\begin{align}
(L_\uu \cc)_{_\UU} = &
(\partial_0 + \om + \vv_{_\UU}\cdot\nnab)\cc_{_\UU} -
\cc_{_\UU}\cdot((\nnab\vv_{_\UU})^* - \om) = \nonumber\\
=& (\partial_0  + \vv_{_\UU}\cdot\nnab)\cc_{_\UU} -
\cc_{_\UU}\cdot(\nnab\vv_{_\UU})^*
\label{uLderv}\end{align}
which is exactly the known form of the {\em upper convected time
derivative}.

Thus, the upper convected time derivative of a spacelike vector
field is just its Lie derivative by the velocity field of the
continuum.

\subsubsection{Lower convected time derivatives}

 An argument similar to that in the previous Subsection
yields that, instead of $\ts\mapsto \K(\Upsilon_\ts(x))$, it seems
preferable to consider $\ts\mapsto (\D
\Upsilon_\ts(x))^*\K(\Upsilon_\ts(x))$ as the change in time of the
covector field $\K:M\to\M^*$ along a particle of the continuum. Then
we find
\begin{equation}\label{colie}
\left.\frac{\dd(\D \Upsilon_\ts(x))^*\K(\Upsilon_\ts(x))}{\dd\ts}
\right|_{\ts=0} = \uu(x)(\D\K)^*(x)+(\D\uu)^*(x)\K(x) =:(L_\uu\K)(x)
\end{equation}

We mention that $L_\uu\K$ is known
in differential geometry as the Lie derivative of $\K$ by $\uu$.

The first term of the Lie derivative is just the material time
derivative.

Recall that $(\D\uu)^*(x)$ is in $\M^*\otimes\frac{\E}{\I}$,
therefore the second term can be written in the form
$(\D\uu)^*(x)\K(x)\cdot\ii$ where $\K(x)\cdot\ii$ is the absolute
spacelike part of the covector field. As a consequence, taking the
absolute spacelike part of \eqref{colie} and putting
$\kk:=\K\cdot\ii$ for the sake of brevity, we have
\begin{equation}\label{Ldercv}
(L_\uu\kk)\cdot\ii= \D_\uu\kk + (\nabla\uu)^*\kk.
\end{equation}

The $\UU$-relative form of the spacelike part of the Lie derivative
of $\kk$ is
\begin{align}\label{uLdercv}
\bigl(L_\uu \kk)\cdot\ii\bigr)_{_\UU} = &
(\partial_0 + \om + \vv_{_\UU}\cdot\nnab)\kk_{_\UU} +
((\nnab\vv_{_\UU})^* - \om)\kk_{_\UU} = \\
=& (\partial_0  + \vv_{_\UU}\cdot\nnab)\kk_{_\UU} +
(\nnab\vv_{_\UU})^*\cdot\kk_{_\UU}
\end{align}
which is exactly the known form of the {\em lower convected time
derivative}.

Thus, the lower convected time derivative of a the spacelike part of a
covector field is just its Lie derivative by the velocity field
of the continuum.

\subsubsection{Jaumann derivative}

According to the identification $\E^*\equiv\frac{\E}
{\mathbf D\otimes\mathbf D}$, a spacelike vector field can be
considered as a covector field and vice versa. Thus, we
can form both the lower convected time derivative and the
upper convected time derivative of a spacelike vector field
$\cc$. In this way we obtain the {\em Jaumann derivative}:
\begin{equation}\label{jau}
J_\uu\cc:= \frac{1}{2}\left(L_\uu\cc + (L_\uu\cc^*)\cdot\ii\right) =
\D_\uu\cc + \frac{\nabla\land\uu}{2}\cc
\end{equation}
whose relative form, according to an observer $\UU$, is
\begin{equation*}
\left(J_\uu\cc\right)_{_{\UU}}=
(\partial_0  + \vv_{_\UU}\cdot\nnab)\cc_{_\UU} +
\frac{\nnab\land \vv_{_\UU}}{2}\cc_{_\UU}.
\end{equation*}

The Jaumann derivative, alternatively, is called the `corotational
time derivative' because it is usually  stated that the Jaumann
derivative is the time derivative with respect to an observer
corotating with the continuum.

The Jaumann derivative, however, is an absolute object,
i.e. independent of any observers, so we have to give a sense
to the above statement, if possible.

First of all note, that a rigid observer cannot corotate totally
with the continuum because the angular velocity of a rigid observer
depends only on time (is the same for all simultaneous world points)
whereas the angular velocity of a (non-rigid) continuum depends on
space-time (is different, in general, for simultaneous space-time
points).

Choosing a single particle of the continuum, we can define a rigid
observer corotating with the continuum around this particle only, in
other words, the space origin of the observer is that particle and
it has angular velocity equalling the angular velocity of the
continuum at that particle. More closely, choosing a single particle
of the continuum described by the world line function $t\mapsto
r_o(t):=\Upsilon_{t-\tau(o)}(o)$ for a given world
point $o$, we put
\begin{equation}\label{joom}
\Om_o(t):=-\frac{(\nabla\land\uu)(r_o(t))}{2}
\end{equation}
and the rigid observer corotating with the continuum around the chosen
particle will be
\begin{equation}
\UU_o(x):=\uu(r_o(\tau(x))) + \Om_o(\tau(x))(x-r_o(\tau(x)).
\end{equation}

The rotation of this observer is obtained as the solution of the
differential equation $\dot \R_o=\Om_o\R_o$ with the initial value
$\R(\tau(o))=\mathrm{id}_\E$.

Then for a spacelike vector field $\cc$ we find by \eqref{slvecsp} that
\begin{equation*}
\partial_0\cc_{_{\UU_o}}= -\R_o^{-1}\dot\R_o\R_o^{-1}\cc(P_o) +
\R_o^{-1}(\D\cc)(P_o)\cdot\UU(P_o)= -(\Om_o\cc)_{_{\UU_o}} +
(\D_{\UU_o}\cc)_{_{\UU_o}}
\end{equation*}
and (see \eqref{jau})
\begin{equation}
\left(J_\uu\cc\right)_{_{\UU_o}}=
(\D_\uu\cc)_{_{\UU_o}} + \left(\frac{\nabla\land\uu}{2}\cc\right)_{_{\UU_o}}.
\end{equation}

According to our choice, $P_o(t,\boldsymbol 0)=r_o(t)$ (see
\eqref{invsplit}), thus \eqref{joom} and $\UU_o(r_o(t))=\uu(r_o(t))$
result in
\begin{equation}
\partial_0\cc_{_{\UU_o}}(t,\boldsymbol 0) =
\left(J_\uu\cc\right)_{_{\UU_o}}(t,\boldsymbol 0):
\end{equation}

-- \ the partial time derivative of the $\UU_o$-relative form

-- \ the $\UU_o$-relative form of the Jaumann derivative
\smallskip

\noindent of a spacelike vector field are equal at the given particle around
which the observer corotates with the continuum.

\section{Relative forms of Lie derivatives}

In the previous sections we have given upper and lower convected derivatives
as relative forms of Lie derivatives of spacelike vectors and covectors.
Further considerations in continuum physics require Lie derivatives
of non-spacelike vectors, covectors and various tensors as well; they
will be treated in this section in a concise form.
Relative forms will be given, too. As sometimes the notation of dual
fields and transposes can be confusing we give the final formulas
also with indexes, for the sake of easier readability. The
$\UU$-relative forms are defined by the splittings \re{hder} and
\re{dhp} for the contravariant and covariant components of the
fields respectively as it was show in section \re{splisec}.

For example the $\UU$-relative velocity field \re{urelu} of the
continuum is written with $\UU$-relative quantities and with an
indexed form, as
$$
 \uu_{_\UU} = \begin{pmatrix} 1 \\ \vv_{_\UU}\end{pmatrix},
 \qquad u^\alpha =\begin{pmatrix} u^0 \\ u^i\end{pmatrix}
$$

\noindent where $u^0=1$. The Greek indexes are denoting four-vectors
and the Roman indexes three vectors, e.g. $\alpha = \{0,1,2,3\}$ and
$i\in\{1,2,3\}$. We should keep in mind that although the indexed
formulas are providing a simple algorithmic method of the
calculations, they are always referring to relative quantities.

\subsection{Scalar fields}

The $\UU$-relative form of a scalar field $f:M\to\rr$ was defined in
\re{scalsp} as $f_{_\UU}=f(P)$. The Lie derivative of $f$
corresponds to its material time derivative and the $\UU$-relative
form of the material time derivative corresponds to the substantial
derivative of the $\UU$-relative form of the scalar field
 \bea
 L_\uu f &=& \left(\left. \frac{d}{dt}
    f(\Upsilon_\ts)\right|_{\ts=0}\right)_\UU =
 (D_\uu f)_{_\UU} = (\D f)_{_\UU}\cdot \uu_{_\UU} \nonumber \\
  &=& (\partial_0 f, {\boldsymbol \nabla}f)
    \bma 1 \\ {\bf v}_{_\UU} \ema =
 \partial_0 f + {\bf v}_{_\UU}\cdot{\boldsymbol \nabla}f.
\eea

With and index notation we can write
$$
 \dot f := u^\alpha\nabla_\alpha f  = \partial_0 f +
 u^i\partial_i f.
$$

\noindent where the dot denotes the substantial derivative.

\subsection{Vector fields}

According to \re{vecsp}, the $\UU$-relative form of a \emph{vector
field} $\C:M\to\M$ is given as
$$
 \C_{_\UU} = \D H(P)\C(P) =: \bma c^0 \\ \cc \ema.
$$

Here we have introduced a convenient notation for the timelike and
spacelike components, $c^0$ and $\cc$ respectively. Therefore, the
Lie derivative of the vector field according to \re{Lvect} is
\ben
 (L_\uu \C)_{_\UU} = \left(\left. \frac{d}{dt} (\D\Upsilon_\ts)^{-1}
    \C(\Upsilon_\ts)\right|_{\ts=0}\right)_\UU =
 (D \C)_{_\UU} u_{_\UU} - (D\uu)_{_\UU} \C_{_\UU}.
\een

Moreover, for rigid observers the corresponding $\UU$-relative form
of the derivatives, can be calculated by partial derivation in the
objective combination, as in \re{uLderv}. The angular velocity of
the observer (the Christoffel symbols) do not play a role. Therefore
the result of the calculations can be written in a simple form
 \bea
 (L_\uu \C)_{_\UU} &=&
 (\partial_\uu \C_{_\UU}) u_{_\UU} - (\partial_\uu u_{_\UU})
 \C_{_\UU}=
 \nonumber\\
  &=& \bma \partial_0 c^0 & {\boldsymbol \nabla}c^0 \\
    \partial_0 \cc & {\boldsymbol \nabla}\cc \ema
    \bma 1 \\ \vv_{_\UU}\ema -
    \bma 0 & {\bf 0} \\
    \partial_0 \vv_{_\UU} & {\boldsymbol \nabla}\vv_{_\UU} \ema
    \bma c^0 \\ \cc \ema = \nonumber\\
  &=& \bma \dot c^0 \\ \dot\cc -
    \cc\cdot{\boldsymbol\nabla}\vv_{_\UU} -
    c^0\partial_0\vv_{_\UU}\ema.
\eea

Hence,
 \bean
 (L_\uu C)^\alpha &=& u^\beta\partial_\beta C^\alpha -
    C^\beta\partial_\beta u^\alpha = \bma
 \partial_0 c^0 + u^j\partial_j c^0 \\
 \partial_0 c^i + u^j\partial_j c^i -
    c^j\partial_j u^i - c^0\partial_0 u^i \ema \\
    &=& \bma \dot c^0 \\
 \dot c^i- c^j\partial_j u^i - c^0\partial_0 u^i \ema
\eean

In the special case when the vector field is the velocity field of
the continuum $\C = \uu$, we get that $L_\uu \uu = {\bf 0}$. The
change of the velocity of the continuum is constant when related to
the continuum.

We can repeat our previous results with our simpler notation and get
the upper convected time derivative as a Lie derivative of a
\emph{spacelike vector field} $\cc:M\to \E$ as \re{uLderv}
substituting the $c^0 = 0$ condition into the formulas above
 \ben
 L_\uu \cc = \bma 0 \\ \dot\cc -
    \cc\cdot({\boldsymbol\nabla}\vv_{_\UU})^* \ema,
\een

\noindent that is
 \ben
 (L_\uu C)^\alpha =  \bma
 0 \\ \dot c^i- c^j\partial_j u^i \ema.
\een

Here we did not omit the trivial zero component. Let us emphasize
again, the results above are written in a form that is similar to
the usual indexed notation, however, the meaning of the symbols are
different and the results are observer dependent. E.g. $\partial_0$
corresponds to the usual partial time derivative and
${\boldsymbol\nabla}$ to the usual coordinatized spacelike
derivations only in case of inertial observers. From the observer
independent, absolute forms one can always calculate the particular
observer dependent splittings as we have seen for rigid observers
above.

\subsection{Covector fields}

We introduce the following notation for the $\UU$-relative form of a
\emph{covector field} $\K:M\to\M^*$, according to \re{covecsp}, as
$$
 \K_{_\UU} = ((\D H(P))^{-1})^*\K(P) =: (k_0, \kk).
$$
The $\UU$-relative form the Lie derivative is \re{colie}
\bea
 (L_\uu \K)_{_\UU} &=& \left(\left. \frac{d}{dt} (\D\Upsilon_\ts)^*
    \K(\Upsilon_\ts)\right|_{\ts=0}\right)_\UU =
 (D_\uu \K_{_\UU}) \uu_{_\UU} + (D_\uu \uu_{_\UU})^* \K_{_\UU}\nonumber\\
  &=& \bma \partial_0 k_0 & {\boldsymbol \nabla}k_0 \\
    \partial_0 \kk & {\boldsymbol \nabla}\kk \ema
    \bma 1 \\ \vv_{_\UU}\ema +
    (k_0, \kk)\bma 0 & {\bf 0} \nonumber\\
    \partial_0 \vv_{_\UU} & {\boldsymbol \nabla}\vv_{_\UU} \ema
      \\
  &=& (\dot k_0 + \kk\cdot\partial_0 \vv_{_\UU},\quad \dot\kk +
    \kk\cdot({\boldsymbol\nabla}\vv_{_\UU}).
\eea

Hence,
 \bean
 (L_\uu K)_\alpha &=& u^\beta\partial_\beta K_\alpha +
    K_\beta\partial_\alpha u^\beta \\
 &=& (\partial_0 k_0 + u^j\partial_j k_0 +k_j\partial_0 u^j, \quad
    \partial_0 k_i + u^j\partial_j k_i +
    k_j\partial_i u^j)\\
 &=& (\dot k_0 + k_j\partial_0 u^j, \dot k_i + k_j\partial_i u^j).
 \eean

One can get the Lie derivative of a \emph{spacelike covector field}
$\kk:M\to \E^*$ substituting $k^0 = 0$ into the formulas above
 \ben
 L_\uu \kk = (\kk\cdot\partial_0 \vv_{_\UU}, \quad \dot\kk +
    \kk\cdot({\boldsymbol\nabla}\vv_{_\UU}),
\een

\noindent that is
 \ben
 (L_\uu k)_i =  (k_j\partial_0 u^j, \quad \dot k_i +  k_j\partial_i u^j).
 \een

Therefore one can see, that {\em the $\UU$-relative form of the Lie
derivative of a spacelike covector field in not spacelike}. The
lower convected time derivative is the spacelike component of the
Lie derivative of a spacelike covector field.

\subsection{Second order tensor fields}

Similarly, the $\UU$-relative form of a \emph{tensor field}
${\mathbf T} :M\to \M\otimes\M$ can be written as
$$
 \T_{_\UU} = \D H(P)\D H(P)\T(P) =: \bma t^{00} & \ts^a \\
 \ts^b & \ol{\bf t}  \ema \in (\I\times \E)\otimes(\I\times\E).
$$

The components of $\T_{_\UU}$ can be calculated according to the
definition of the observer splittings \re{vecsp}, one can apply
\re{hder} to both components of the tensorial product. We may
recognize, that only the time-timelike component of the second order
contravariant tensor is independent on the observer $t^{00} =
{\boldsymbol \tau} {\boldsymbol \tau}\T$. The $\UU$-relative form of
the Lie derivative of $\T$ expressed by the relative quantities is
 \begin{gather}
 (L_\uu \T)_{_\UU} = \left(\left. \frac{d}{dt}
 (\D\Upsilon_\ts)^{-1}(\D\Upsilon_\ts)^{-1}
    \T(\Upsilon_\ts)\right|_{\ts=0}\right)_\UU =\nonumber\\
    =
 (D_\uu \T_{_\UU}) \uu_{_\UU} -  (D_\uu \uu_{_\UU}) \T_{_\UU}
    - \T_{_\UU}(D_\uu \uu_{_\UU})^* =\nonumber\\
  = \bma \dot t^{00} &
    \dot \ts^a - t^{00} \partial_0 \vv_{_\UU} -
        \ts^a \cdot\nabla \vv_{_\UU} \\
    \dot \ts^b - t^{00} \partial_0 \vv_{_\UU} -
        \ts^b \cdot\nabla \vv_{_\UU} &
    \dot{\ol{\bf t}} - \partial_0 \vv_{_\UU}\ts^a -
        \ts^b \partial_0 \vv_{_\UU} -
        \ol{\bf t} \cdot(\nabla \vv_{_\UU})^* -
        (\nabla \vv_{_\UU})\cdot\ol{\bf t}
       \ema.
\end{gather}

With the indexed notation we get
 \begin{align}
 &(L_\uu T)^{\alpha\beta} =
    u^\gamma\partial_\gamma t^{\alpha\beta} -
    t^{\gamma\beta}\partial_\gamma u^\alpha -
    t^{\alpha\gamma}\partial_\gamma u^\beta = \\
  &  = \bma
        \dot t^{00} &
        \dot t^{0j} - t^{00}\partial_0 u^j - t^{0k}\partial_k u^j \\
        \dot t^{i0} - t^{00}\partial_0 u^i - t^{k0}\partial_k u^i &
        \dot t^{ij} - t^{0j}\partial_0 u^i - t^{kj}\partial_k u^i -
            t^{i0}\partial_0 u^j - t^{ik}\partial_k u^j
    \ema .\nonumber
\end{align}

 If $\T$ is space-spacelike, we substitute $t^{00}=0$, $\ts^a =
{\bf 0}$ and $\ts^b ={\bf 0}$ into the previous formula. The
$\UU$-relative form of the Lie derivative of a space-spacelike
second order tensor is space-spacelike and we can get the upper
convected derivative of the three dimensional second order tensor.
 \bea
 L_\uu \bma 0 & 0 \\ 0 & \ol{\bf t} \ema =
 \bma
    0 & 0\\ 0 & \dot{\ol{\bf t}} -\ol{\bf t}\cdot(\nabla \vv_{_\UU})^* -
        (\nabla \vv_{_\UU})\cdot\ol{\bf t}
 \ema.
\label{spst}\eea

\subsection{Second order cotensor fields}

The $\UU$-relative form of a \emph{cotensor field} ${\mathbf W}
:M\to \M^*\otimes\M^*$ can be written as
$$
 \W_{_\UU} = (\D H(P)^{-1})^*(\D H(P)^{-1})^*\W(P) =:
    \bma w_{00} & \w_a \\
 \w_b & \ul{\bf w}  \ema \in (\I^*\times \E^*)\otimes(\I^*\times\E^*).
$$

The detailed form of $\W_{_\UU}$ can be calculated according to the
definition of the observer splittings \re{dhp}. The $\UU$-relative
form of the Lie derivative of $\W$ expressed by the relative
quantities is
 \bea
 (L_\uu \W)_{_\UU} &=& \left(\left. \frac{d}{dt}
 \D\Upsilon_\ts\D\Upsilon_\ts
    \W(\Upsilon_\ts)\right|_{\ts=0}\right)_\UU  =\\
    &=& (D_\uu \W_{_\UU}) \uu_{_\UU} + (D_\uu \uu_{_\UU})^* \W_{_\UU}
    + \W_{_\UU}(D_\uu \uu_{_\UU}) = \nonumber\\
  &=& \bma \dot w_{00} + \partial_0\vv_{_\UU}(\w_a+\w_b) &
    \dot \w_a + \nabla \vv_{_\UU}\cdot\w_a +
        \partial_0 \vv_{_\UU}\cdot\ul{\bf w} \\
    \dot \w_b + \nabla \vv_{_\UU}\cdot\w_b +
        \ul{\bf w}\cdot\partial_0 \vv_{_\UU} &
     \dot{\ul{\bf w}} +
        \ul{\bf w} \cdot(\nabla \vv_{_\UU}) +
        (\nabla \vv_{_\UU})^*\cdot\ul{\bf w}
       \ema.
\nonumber\eea

With the indexed notation we get
 \bean
 (L_\uu W)_{\alpha\beta} &=&
    u^\gamma\partial_\gamma W_{\alpha\beta} +
    W_{\gamma\beta}\partial_\alpha u^\gamma  +
    W_{\alpha\gamma}\partial_\beta u^\gamma \nonumber\\
    &=& \bma
        \dot w_{00} + w_{k0}\partial_0 u^k + w_{0k}\partial_0 u^k &
        \dot w_{0j} + w_{kj}\partial_0 u^k + w_{0k}\partial_j u^k \\
        \dot w_{i0} + w_{k0}\partial_i u^k + w_{ik}\partial_0 u^k &
        \dot w_{ij} + w_{kj}\partial_i u^k + w_{ik}\partial_j u^k
    \ema .
\eean

If $\W$ is space-spacelike, we can substitute $w_{00}={\bf 0}$,
$\w_a = {\bf 0}$ and $\w_b =0$ into the previous formula. {\em The
$\UU$-relative form of the Lie derivative of a space-spacelike
second order cotensor is not space-spacelike}. Its space-spacelike
component is the lower convected derivative of the space-spacelike
component of $\W$.
 \bea\label{spsct}
 (L_\uu \ul{\bf w})_{_{\uu}} = L_\uu \bma 0 & 0 \\
    0 & \ul{\bf w} \ema =
 \bma
     0 & \ul{\bf w}\cdot \partial_0\vv_{_\UU} \\
    \ul{\bf w}\cdot\partial_0 \vv_{_\UU} & \dot{\ul{\bf w}} +
        (\nabla \vv_{_\UU})^* \cdot\ul{\bf w}
        + \ul{\bf w}\cdot(\nabla \vv_{_\UU})
 \ema.
\eea

\subsection{Second order mixed tensor fields}

The $\UU$-relative form of a \emph{mixed field} ${\mathbf A} :M\to
\M\otimes\M^*$ can be written as
$$
 \A_{_\UU} = ((\D H)(P))((\D H(P))^{-1})^*\A(P) =:
    \bma A^0_0 & {\mathbf a}_a \\
 {\mathbf a}^b & \ol{\ul{\bf a}}  \ema
 \in (\I\times \E)\otimes(\I^*\times\E^*).
$$

The detailed form of $\A_{_\UU}$ can be calculated according to the
definitions of the observer splittings \re{hder} and \re{dhp}. The
$\UU$-relative form of the Lie derivative of $\A$ can be expressed
by the relative quantities as
 \begin{gather}\label{mtLsp}
 (L_\uu \A)_{_\UU} = \left(\left. \frac{d}{dt}
 (\D\Upsilon_\ts)^{-1}\D\Upsilon_\ts
    \A(\Upsilon_\ts)\right|_{\ts=0}\right)_\UU \\
 = (D_\uu \A_{_\UU}) \uu_{_\UU} - (D_\uu \uu_{_\UU}) \A_{_\UU}
    + \A_{_\UU}(D_\uu \uu_{_\UU}) =\nonumber\\
 = \bma \dot A^0_0 + {\mathbf a}_a\cdot\partial_0\vv_{_\UU} &
    \dot {\mathbf a}_a + {\mathbf a}_a \cdot \nabla \vv_{_\UU}=
    \nonumber\\
    \dot {\mathbf a}^b - A^0_0\partial_0 \vv_{_\UU}-
        {\mathbf a}^b\cdot \nabla \vv_{_\UU} +
        \ul{\ol{\bf a}}\cdot\partial_0 \vv_{_\UU} &
     \dot{\ul{\ol{\bf a}}} - \partial_0\vv_{_\UU}\cdot {\mathbf a}^b +
        \ul{\ol{\bf a}} \cdot(\nabla \vv_{_\UU}) -
        (\nabla \vv_{_\UU})\cdot\ul{\ol{\bf a}}
       \ema.
\nonumber\end{gather}

With the indexed notation we get
 \begin{gather}
 (L_\uu A)^\alpha_\beta =
    u^\gamma\partial_\gamma A^\alpha_\beta -
    \partial_\gamma u^\alpha A^\gamma_\beta  +
    \partial_\beta u^\gamma A^\alpha_\gamma =\\
    = \bma
        \dot a^0_0 + a^0_k\partial_0 u^k  &
        \dot a^0_j + a^0_k\partial_j u^k \\
        \dot a^i_0 - a^0_0\partial_0 u^i - a^k_0\partial_k u^i
            + a^i_k\partial_0 u^k &
        \dot a^i_j - a^0_j\partial_0 u^i - a^k_j\partial_k u^i
            + a^i_k\partial_j u^k
    \ema . \nonumber
\end{gather}

If $\A$ is space-spacelike, we can get the proper formula by
substituting $a^0_0=0$, ${\bf a}_a = {\bf 0}$ and ${\bf a}^b ={\bf
0}$ into \re{mtLsp}. {\em The $\UU$-relative form of the Lie
derivative of a space-spacelike second order cotensor is not
space-spacelike} in general.
 \bea\label{spsmt}
 (L_\uu \ul{\ol{\bf a}})_{_{\uu}} = L_\uu \bma 0 & 0 \\
    0 & \ul{\ol{\bf a}} \ema =
 \bma
        0  & 0 \\
        a^i_k\partial_0 u^k &
        \dot a^i_j - a^k_j\partial_k u^i + a^i_k\partial_j u^k
 \ema.
\eea

\section{Summary}

In this paper we investigated objective time derivatives of
continuum physics in a four-dimensional setting. Our analysis was based on a
reference frame independent nonrelativistic space-time model in which time is
not embedded into space-time.

Within this space-time model a definition of objectivity (frame
independence) was introduced by the use of absolute objects -- four-vectors,
covectors, tensors etc. -- not referring to any observer.
Of course, observers are defined in  this theory, and detailed formulae are
given, how space-time is split into time and space by a rotating observer and
how absolute objects are split into time- and spacelike components.

Considering continuous media, we have defined material time
differentiation in an absolute form (not depending on observers). Its
correct relative form corresponding to a rotating observer $\UU$ is
the substantial differentiation
$\partial_0+{\bf v}_{\UU}\cdot{\mathbf \nabla}$ only for scalars;
for spacelike vectors it is $\partial_0+ {\mathbf \Omega}_\UU + {\bf
v}_{\bf U}\cdot{\mathbf \nabla}$.

The four-dimensional Lie derivatives of scalars, vectors, covectors,
second order tensors, cotensors and mixed tensors were calculated
together with their usual relative forms. For the calculation we have
introduced a simplified formalism exploiting that in the Lie
derivatives the Christoffel symbols of the coordinatization do not
appear. We have found that in some cases the Lie derivatives
correspond to well known objective time derivatives. For example,
the Lie derivative of spacelike vectors is the upper convected time
derivative, the spacelike component of the Lie derivative of
covectors is the lower convected time derivative, etc... The four-dimensional
treatment was essential because the Lie derivative of
covectors is not spacelike in general.

\section{Discussion}

Notions of differential geometry (as e.g. Lie derivatives or
Christoffel symbols) are tools of formulating the general principles
of continuum mechanics \cite{MarHug83b} and are also important in
modeling the microstructure \cite{FabMar05a,Sac01a}. In most of the
related investigations the geometry is related only to the three
dimensional space, as the time dependence is introduced in a trivial
way, space-time is considered to be the Cartesian product of space and time.
Some recent treatments introduce non-relativistic space-time with
geometrical notions, as a simple fibre bundle \cite{SveBer99a}. The
space-time model of our paper is the simplest possible one, with the
same structure. However, our
researches show, that the four-dimensional structure cannot be
avoided with any reference to e.g. 'instantanous transformations'.
The few existing four-dimensional treatments (e.g.
\cite{TruTou60b,Mau00a,KieHer03a}) do not consider the problem of objectivity
and objective time derivatives. In a previous paper we
have argued, that objectivity cannot be formulated properly in three
dimension, because the proper transformation of physical quantities
between time dependent (e.g. rotating) reference frames require the
use of four-dimensional Christoffel symbols \cite{MatVan06a}.

Objective time derivatives appear mostly in rheology. As we have
already mentioned in the introduction, their construction is
originally based on an ad-hoc supplementation of the substantial
time derivative \cite{Jau11a,BamMor80a}. Contrary to the fact that
the best phenomenological models of rheology contain objective time
derivatives, an extensive experimental research showed that their
applicability is restricted and the comparison of the different
rheological models demonstrate essential differences
\cite{BirAta77b,Tan85b}. For example a model can give good
viscometric functions in case of simple shear, but fail to explain
results of other experiments related to the very same material. Let
us remark that in a usual rheological model the same objective time
derivative is used for physical quantities of
different tensorial character (contrary to the well known facts that
stress is a second order tensor,
strain is a second order mixed tensor).

That later point deserves closer attention, because according to our
investigations the objective time derivatives of a physical quantity
can be different and depend on its tensorial properties. In three
dimension the distinction of vectors and covectors implicitly
appears already in the original works of Oldroyd \cite{Old49a}
introducing convected derivatives, later also by Lodge
\cite{Lod74b}. In the basic books of rheology \cite{BirAta77b} and
later developments as finite strain viscoelasticity \cite{ReeGov98a}
the use of Lie derivatives (convected time derivatives) is a
standard. However, the careful distinction of vectors and covectors
is rare and is restricted to three dimensions. For example in the
investigations of Haupt and his coworkers (see \cite{Hau00b} and the
references therein) upper and lower convected time derivatives are
connected to vectors and covectors, similarly as in our
intrinsically four dimensional investigations. However, as we have
pointed out in section \re{st}, introducing the space-time model, in
case of spacelike quantities there is a way to identify the two
quantities and therefore to transform a vector to a covector and
back. Haupt and coworkers make this identification for the stress
and strain rate tensors requiring the invariance of the power.
However, this cannot be a general solution, there are objective time
derivatives of physical quantities of non mechanic origin, without
the kind of physical duality expressed by the power.

Continuum mechanics and rheology are not compatible regarding time
derivatives. In rheology - a mechanic theory of generalized fluids -
objective time derivatives are unavoidable. In material theories of
modern mechanics - especially the ones based on the concept of
material manifold (see e.g. \cite{Mau99b,KieHer00b}) - only the
substantial time derivative appear. Experience shows that there is
no need any of the objective derivatives of the deformation gradient
${\boldsymbol \chi}_{\bf t}$ in finite strain mechanics with or
without memory. However, in our frame the objective time derivative
of any second order tensor is seemingly different of the material
time derivative \re{spst}, \re{spsct}, \re{spsmt}. This apparent
contradiction can be easily explained recognizing that the motion-related
physical quantities can have special Lie derivatives. E.g.
the deformation gradient field  ${\bf F}_{\bf t} =\nabla\chi_{\bf
t}(\Xx)$ is the $\UU$-relative spacelike form of the mixed four
tensor field $D\Upsilon_{\bf t}$. Therefore substituting it into the
definition of the Lie derivative \re{mtLsp} we  get
\begin{gather}
 (L_\uu \A)_{_\UU} = \left(\left. \frac{d (\D\Upsilon_\ts)^{-1}\D\Upsilon_\ts
    \D\Upsilon_\ts}{dt}(\Upsilon_\ts)
 \right|_{\ts=0}\right)_\UU =
    \nonumber\\=
    \left(\left. \frac{d}{dt}
    \D\Upsilon_\ts(\Upsilon_\ts)\right|_{\ts=0}\right)_\UU =
    (D_\uu \uu_{_\UU})(\D\Upsilon_\ts)
 \end{gather}

\noindent for the space-spacelike part we easily get $\dot{\bf
F}_{\bf t} = (\nabla\vv_{_\UU})\cdot{\bf F}_{\bf t}$, giving the
well established relation of the velocity gradient and the material
derivative of the deformation gradient. An other important motion-related
physical quantity is the velocity and we have seen that its
Lie derivative is zero. That can explain why velocity cannot appear
in material functions without referring to the usual form-invariance
arguments.

Our investigations are related to the
principle of material frame indifference through the new, four
dimensional definition of objectivity. The problem regarding the
proper formulation of frame independent material equations still
lacks a generally accepted solution and the related discussions are
not settled \cite{Mur03a,Liu03a,Mur05a,Liu05a}. A clear exposition
of the problem is given e.g. by \cite{EdeMcL73a,Rys85a}. The
inevitable fact is that the traditional phenomenological formulation
of the above mentioned evident requirement \cite{Nol58a,TruNol65b}-
that the constitutive equations characterizing the materials should
be independent on the outer observer - are paradoxically
contradicting the results of the kinetic theory. There are opinions
that kinetic theory is not frame independent
\cite{Mul72a,MulRug98b}, that material frame independence is only an
approximations \cite{Spe98a} and that frame independence should be
redefined on the phenomenological level. One of the related attempt
exploits the objectivity of the balance form equations of continuum
physics \cite{Mus98a}. An important step in this respect is the
careful distinction of the related principles and concepts is given
by Svensen and Bertram \cite{SveBer99a,BerSve01a}. According to
their investigations there are three related concepts in this
respect: Euclidean frame-indifference (objectivity), form invariance
of the material functions and indifference with respect to
superimposed rigid body motions. They have shown that the validity
of any two of these concepts automatically imply the third. Our
general opinion is that a four-dimensional, space-time formulation
of objectivity is unavoidable. That could explain both the results
of kinetic theory \cite{Mat86a,MatGru96a} and shows clearly that in
the original definition of objectivity only a spacelike part of a
space-time transformation was considered and four-dimensional
Christoffel symbols were neglected \cite{MatVan06a}. In this case
the cited implication of Svendsen and Bertram requires further
investigations.

Finally let us point out three fields where we think that the
consequences of our basic mathematical investigation can be checked
and can lead to further understanding of new physical phenomena and
formulation of new theories of continuum physics
\begin{itemize}
    \item A proper objective time derivative depends on the
    tensorial properties of the physical quantity. Objective time
    derivatives of vectors and covectors are different. As a
    consequence one can expect that different physical quantities
    (e.g. the mixed strain tensor and the stress cotensor)
    can have different objective time derivatives
    in the very same rheological model.
    Let us remark that for a good model construction a constructive
    thermodynamic theory could be essential (See e.g. the simple
    and instructive thermodynamic generalization of the corotational
    Jeffreys model by Verh\'as \cite{Ver97b}.
    \item The objective time derivatives of a spacelike physical
    quantity is not necessarily spacelike. Four-dimensional
    contributions and terms can be important. A good example
    can be, that the four-dimensional GENERIC structure can lead to the concept
    of conductive mass current \cite{Bre05a,Bre05a1,Ott05b}.
    \item Relativistic material theories beyond Newtonian fluids
    \cite{BorChr03a,BorChr05m,HerAta00a} cannot be developed without a
    true definition
    of objectivity. The generalization of our definition in case
    of relativistic space-time models is straightforward.
   \end{itemize}

\section{Acknowledgement}

This research was supported by OTKA T048489 and by a Bolyai
scholarship for P. V\'an.

\bibliographystyle{unsrt}

\end{document}